# FRAME-LEVEL QUALITY AND MEMORY TRAFFIC ALLOCATION FOR LOSSY EMBEDDED COMPRESSION IN VIDEO CODEC SYSTEMS

Li Guo, Dajiang Zhou, Shinji Kimura, and Satoshi Goto

Graduate School of Information, Production and Systems, Waseda University, Japan
Email: guoli@toki.waseda.jp

## ABSTRACT

For mobile video codecs, the huge energy dissipation for external memory traffic is a critical challenge under the battery power constraint. Lossy embedded compression (EC), as a solution to this challenge, is considered in this paper. While previous studies in EC mostly focused on compression algorithms at the block level, this work, to the best of our knowledge, is the first one that addresses the allocation of video quality and memory traffic at the frame level. For lossy EC, a main difficulty of its application lies in the error propagation from quality degradation of reference frames. Instinctively, it is preferred to perform more lossy EC in non-reference frames to minimize the quality loss. The analysis and experiments in this paper, however, will show lossy EC should actually be distributed to more frames. Correspondingly, for hierarchical-B GOPs, we developed an efficient allocation that outperforms the non-reference-only allocation by up to 4.5 dB in PSNR. In comparison, the proposed allocation also delivers more consistent quality between frames by having lower PSNR fluctuation.

## 1. INTRODUCTION

In video codec systems, embedded compression (EC) has been widely applied to address the limited memory bandwidth and the power dissipation from memory traffic. With EC, reference frames are compressed before being stored into the external memory (DRAM), and decompressed after fetched back.

Many EC techniques [1]-[6] have been presented in recent years, which can be classified into two categories of lossless [2][3] and lossy [4]-[6] EC. While lossless EC ensures maintaining the video quality, it is usually regarded being unable to improve the worst-case performance. Lossless EC also inevitably leads to a variable data reduction ratio (DRR) at the block level, which requires the support from an additional address translation mechanism [7]. On the other hand, lossy EC can be based on a fixed DRR which is much easier to implement while contributing to reducing not only the average-case memory traffic but also the requirement for memory bandwidth at the worst case. In the meanwhile, however, lossy EC suffers from quality degradation, especially due to the frame-to-frame error propagation from the loss in the reference frames.

Most of the previous works in lossy EC focused on block-level compression algorithms and architectures that achieve better trade-off between DRR and video quality. The error propagation, however, also highly depends on the group of pictures (GOP) structure and how lossy EC is applied at the frame level. In this work, frame-level quality and memory traffic allocation is analyzed for lossy EC based on the hierarchical-B frame structure. Compared to the low-delay configurations, hierarchical-B has up to 50% proportion of non-reference frames and a much shorter error propagation chain, which makes it a more practical object to apply lossy EC.

For hierarchal-B GOPs, it is an instinctive idea that quality loss can be minimized by eliminating the influence of error propagation, if lossy EC is only performed in the non-reference frames. The analysis and experiments in this paper, however, will demonstrate such an idea is wrong. Apart from the obvious fluctuation of quality from this allocation, which is clearly negative to the visual experience [8], this non-reference-only allocation even performs poorly in terms of average PSNR. In this work, we developed an efficient allocation strategy that distributes EC to both reference and non-reference frames, and achieved up to 4.5 dB PSNR gain. The proposed allocation also delivers a significantly better stability of video quality.

The rest of this paper is organized as follows. Firstly, in Section 2, the relation between video quality degradation and lossy EC is analyzed from two sources of within current frame and between frames. Then from frame level, the frame-level optimal DRR allocation is presented in Section 3. Section 4 shows the experimental results and comparison. Finally, we conclude this paper in Section 5.

## 2. ANALYSIS OF VIDEO QUALITY DEGRADATION FROM LOSSY EC

In a hierarchical-B GOP, video quality degradation is from two sources: lossy compression in the current frame and the error propagation from reference frames. Respectively from these two sources, the relationship between lossy EC used in a video decoder and the corresponding quality degradation is analyzed in this section.

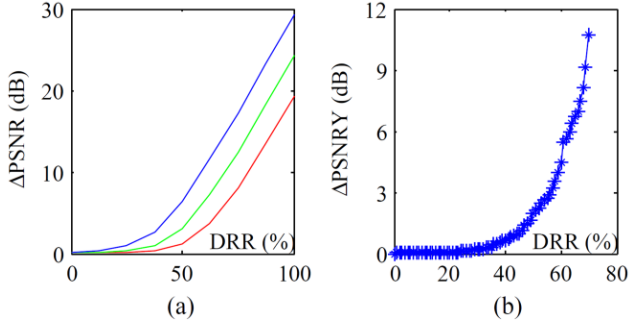

**Fig. 1.** Relationship between compression performance (DRR) and video quality degradation (ΔPSNR). (a) from theoretical analysis [see Eq.6] (Due to the quality loss in video encoding, PSNR$_{w/o}$ is set to 30, 35, 40 dB, which are descripted in red, green, and blue, respectively.), (b) from experimental results.

## 2.1. Lossy EC in the Current Frame

According to the definition of PSNR [see Eq.1], PSNR degradation (ΔPSNR) [see Eq.2] caused by lossy compression is only related to the mean squared error (MSE).

$$PSNR = 10log_{10}\left(\frac{MAX^2}{MSE}\right) \quad (1)$$

$$\Delta PSNR = PSNR_{w/o} - PSNR_{w/} = 10log_{10}\left(\frac{MSE_{w/}}{MSE_{w/o}}\right) \quad (2)$$

where subscripts $_{w/}$ and $_{w/o}$ indicate the cases with and without lossy EC, respectively, during video decoding. For MSE we have:

$$MSE_{w/o} = \frac{1}{N}\sum_{i=1}^{N} e_{i,w/o}^2 \quad (3)$$

$$MSE_{w/} = MSE_{w/o} + \frac{2}{N}\sum_{i=1}^{N} e_{i,w/o}e_{i,ec} + \frac{1}{N}\sum_{i=1}^{N} e_{i,ec}^2 \quad (4)$$
$$= MSE_{w/o} + 2E[e_{w/o}e_{ec}] + E[e_{ec}^2], when\ N \to \infty$$

where $e_{w/o}$ is error purely from video coding, and $e_{ec}$ is the additional error caused by lossy EC in video decoding. Note the plus/minus sign is preserved in $e_{w/o}$ and $e_{ec}$, so the addition of the two can either result in an amplified or cancelled error. *N* is the number of pixels in the frame.

For simplification purpose, linear quantization of the pixel value is considered as the lossy EC approach. In case the *M* least significant bits are truncated, $e_{ec}$ follows a discrete uniform distribution on $[-2^{M-1}, 2^{M-1} - 1]$. Being independent to $e_{ec}$, $e_{w/o}$ also has a zero-mean symmetric distribution. As a result, the expectation of the second component in $MSE_{w/}$ can be decomposed and should be equal to zero:

$$MSE_{w/} = MSE_{w/o} + 2E[e_{w/o}]E[e_{ec}] + E[e_{ec}^2] \quad (5)$$
$$= MSE_{w/o} + \frac{2^{2M-1}+1}{6}$$

Finally, with Eq. 5, ΔPSNR can be expressed as:

$$\Delta PSNR = 10log_{10}\left(\frac{MSE_{w/o} + 4^M/12 + 1/6}{MSE_{w/o}}\right) \quad (6)$$

From the above equation, ΔPSNR is a convex function of variable *M* under a constant $MSE_{w/o}$. We further define DRR as the percentage of the reduced data size from EC. Since DRR is proportional to the reduced number of bits per pixel (*M*). ΔPSNR should be also a convex function of DRR, as plotted in Fig. 1(a). To avoid frame-to-frame error propagation, lossy EC can be performed only in non-reference frames of a hierarchical-B GOP. Under such a configuration, experimental results of a practical lossy EC algorithm showing the relation between DRR and ΔPSNR are plotted as Fig. 1(b), which conform well to our theoretical analysis. Detailed experimental conditions will be given in Section 4.1.

From the convexity, it can be concluded that an even allocation of DRR should achieve better video quality if error propagation is not considered, in comparison to the performing lossy EC only in non-reference frames.

## 2.2. Error Propagation between Frames

### 2.2.1. Theoretical analysis

We consider the reference relation between two frames, where all blocks are inter-coded and EC with the same degree of loss is performed on the current and the reference frames. Compared to the case without lossy EC, there are two additional errors in the current frame: $e_{ec}$ is the error from lossy EC in the current frame; $e_{ep}$ is the propagated error. Specifically, for the *i*-th pixel of the current frame, $e_{i,ep}$ is caused by the EC error of its reference pixel. The resulting MSE of the current frame can be expressed as:

$$MSE_{w/} = \frac{1}{N}\sum_{i=1}^{N}(e_{i.w/o} + e_{i,ec} + e_{i,ep})^2 \quad (7)$$
$$= MSE_{w/o} + 2E[e_{w/o}e_{ec}] + 2E[e_{w/o}e_{ep}]$$
$$+ E\left[(e_{ec} + e_{ep})^2\right], when\ N \to \infty$$

With $e_{w/o}$'s zero-mean distribution and independence to both $e_{ec}$ and $e_{ep}$, components with $e_{w/o}$ can be removed:

$$MSE_{w/} = MSE_{w/o} + 2E[e_{w/o}]E[e_{ecc}] \quad (8)$$
$$+ 2E[e_{w/o}]E[e_{ep}] + E\left[(e_{ec} + e_{ep})^2\right]$$
$$= MSE_{w/o} + E\left[(e_{ec} + e_{ep})^2\right]$$

From Cauchy-Schwarz, we further have the inequality as given in Eq. 9.

$$E\left[(e_{ec} + e_{ep})^2\right] \leq E[e_{ec}^2] + 2E[|e_{ec}e_{ep}|] + E[e_{ep}^2] \quad (9)$$
$$\leq E[e_{ec}^2] + 2\sqrt{E[e_{ec}^2]E[e_{ep}^2]} + E[e_{ep}^2]$$

Since both $e_{ec}$ and $e_{ep}$ follow a discrete uniform distribution on $[-2^{M-1}, 2^{M-1} - 1]$, $MSE_{w/}$ should have an upper bound which is a function of *M*:

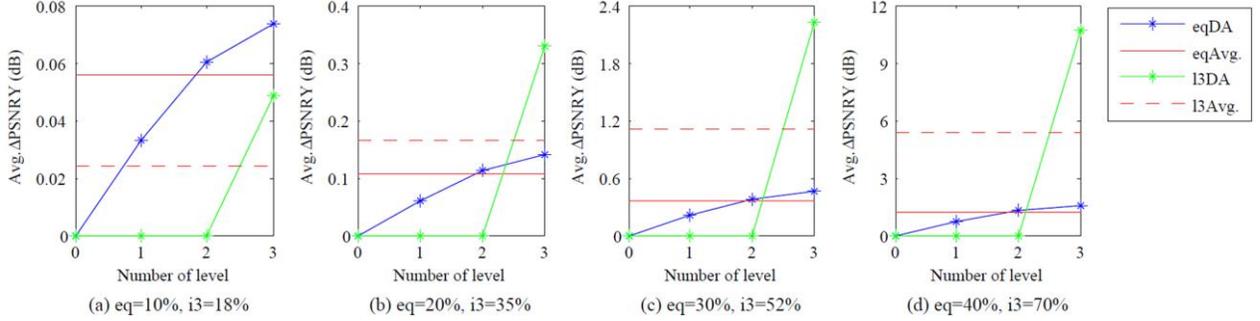

**Fig. 4.** Comparison of average ΔPSNR in each level between evDA and l3DA from experimental results. (a) SeqDRR = 8.75%, (b) SeqDRR = 17.5%, (c) SeqDRR = 26.25%, (d) SeqDRR = 35%. (The red line is average ΔPSNR for all levels.)

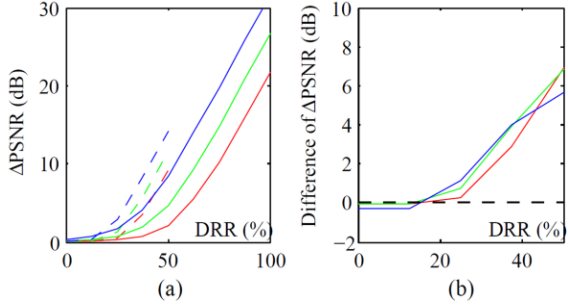

**Fig. 2.** Quality comparison between two allocations of only on non-reference frames and all B frames evenly. (a) ΔPSNR (solid line is for only non-reference frame, and dotted line is allocation evenly), (b) difference of ΔPSNR in (a). (the red, green, and blue lines indicate $PSNR_{w/o,r}$ = 30, 35, 40 dB respectively. $PSNR_{w/o,r}$ - $PSNR_{w/o,c}$ is set to 1 dB. $PSNR_{w/o,r}$ and $PSNR_{w/o,c}$ are video quality of the reference and current frame measured without lossy EC.)

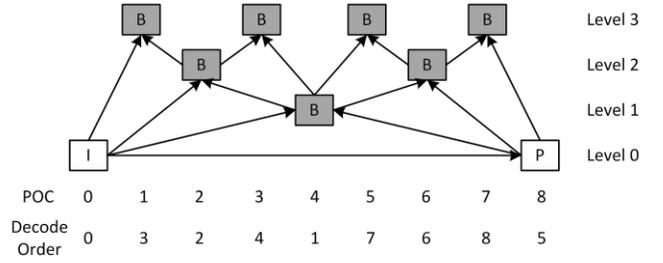

**Fig. 3.** GOP structure and the definition of levels.

$$MSE_{w/} \le MSE_{w/o} + \frac{1}{3}4^M + \frac{2}{3} \quad (10)$$

For ΔPSNR we have:

$$\Delta PSNR \le 10 log_{10}\left(\frac{MSE_{w/o} + 4^M/3 + 2/3}{MSE_{w/o}}\right) \quad (11)$$

As a conclusion, even with error propagation taken into consideration, the upper bound of ΔPSNR is still convex. Based on Eq. 6 and Eq. 11, the upper bound of average ΔPSNR (of the current and reference frames) is plotted in Fig. 2(a) as the solid lines. In video coding, the PSNR of reference frame ($PSNE_{w/o,r}$) is usually larger than the predicted current frame ($PSNE_{w/o,c}$) by 0~2 dB. Therefore in this example, the difference between $PSNE_{w/o,r}$ and $PSNE_{w/o,c}$ is set to 1 dB, while a similar curve and relation is also observed under different configurations.

As a comparison, we analyze two allocations. One of them applies lossy EC only to non-reference frames while the other applies lossy EC to all B frames evenly. For non-reference only allocation, no error will be propagated to other frames. To achieve the same average DRR with the above example of even allocation, non-reference frames need to be truncated by *2M* bits. The difference of average ΔPSNR between these two allocations ($\Delta PSNR_{non-ref}$ and $\Delta PSNR_{even}$) is defined as $\Delta^2 PSNR$ shown in Eq. 12. From Eq. 6 and Eq. 11, its lower bound is obtained as:

$$\Delta^2 PSNR = \Delta PSNR_{non-ref} - \Delta PSNR_{even} \quad (12)$$

$$\ge 10 log_{10}\left[\frac{MSE_{w/o,r} \times \left(MSE_{w/o,c} + \frac{4^{2M}}{12} + \frac{1}{6}\right)}{\left(MSE_{w/o,r} + \frac{4^M}{12} + \frac{1}{6}\right)\left(MSE_{w/o,c} + \frac{4^M}{3} + \frac{2}{3}\right)}\right]$$

where subscripts $_r$ and $_c$ indicate the reference and the current frame, respectively.

When $PSNE_{w/o,r}$ is in the range of 30~40 dB and larger than $PSNE_{w/o,c}$ by 0~2dB, the lower bound of $\Delta^2 PSNR$ is a positive value for $M \ge 2$. In this case, even allocation will achieve smaller ΔPSNR than non-reference only allocation.

To describe this relation clearly, the average ΔPSNR for non-reference only allocation is plotted as the dotted line in Fig. 2(a), and the lower bound of $\Delta^2 PSNR$ is drawn in Fig. 2(b). The negative value indicates the allocation only on non-reference frame achieves smaller ΔPSNR. It is better for even allocation when $\Delta^2 PSNR$ is positive. Hence, to improve video quality, the allocation of lossy EC should be modified from only on the non-reference frames to more B frames evenly with the increase of DRR.

*2.2.2. Analysis from Experimental Results*

As an example, a typical configuration of random access with the GOP size of 8 is show in Fig. 3. According to the reference relation, all hierarchical B frames are divided into three levels (level1/2/3), where lossy EC is performed. I and P frames are viewed as level0. Lossless or non EC is utilized in level0 to avoid error propagation between GOPs. All experiments in this paper will follow this GOP structure. And its sequence-level DRR (SeqDRR) is defined as:

$$SeqDRR = \frac{DRR_{level1}}{8} + \frac{DRR_{level2}}{4} + \frac{DRR_{level3}}{2} \quad (13)$$

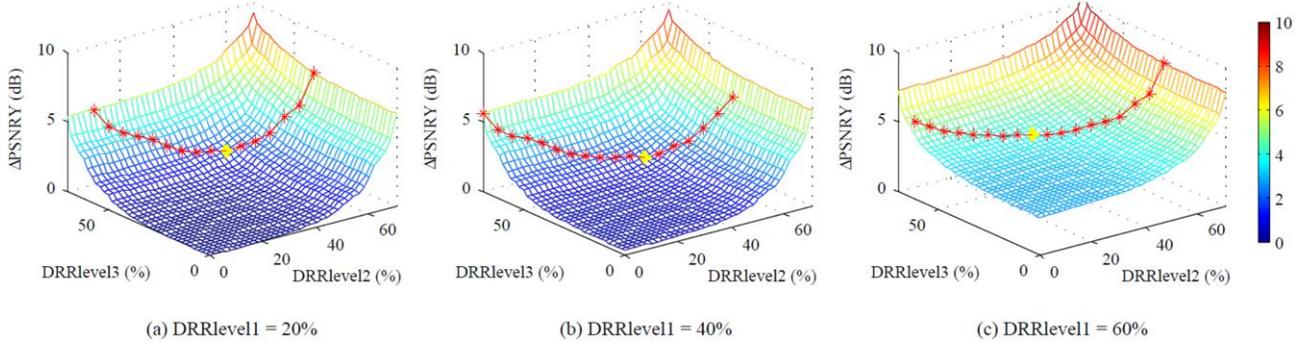

**Fig. 6.** Video quality degradation (ΔPSNR) under different allocation of SeqDRR in level1/2/3 (DRR$_{level1/2/3}$). (a)~(c) ΔPSNR distribution when DRR$_{level1}$ = 20%, 40% and 60% respectively. The red line is isoDRR line for SeqDRR = 40%. The yellow node indicates the minimum ΔPSNR on the isoDRR line (minΔPSNR$_{23}$).

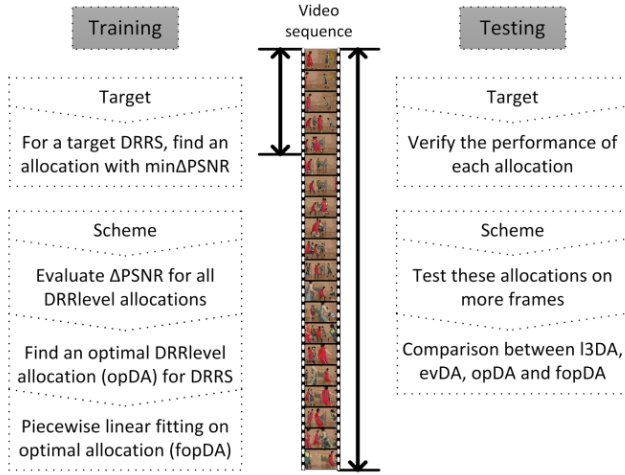

**Fig. 5.** Overall flow for the optimization of DRR allocation.

To evaluate the error propagation, lossy EC with the same fixed DRR is processed on B frames at each level, which is viewed as even DRR allocation (evDA). The blue lines in Fig. 4 show the average ΔPSNR in each level under different SeqDRR. For frames in level1, the only source of ΔPSNR is the lossy compression in the current frame, while the error propagation factor applies to level2 and level3. Experimental results show that, the additional quality loss between two adjacent levels tends to decrease with the accumulation of error propagation.

Under the same SeqDRR, the ΔPSNR for level3 DRR allocation (l3DA) is also shown in Fig. 4 as the green line. l3DA indicates that lossy EC is only performed in level3, so there is no error propagation between frames. Moreover, we calculate the average ΔPSNR of all frames in level0~level3 as the red lines shown.

According to these results, to minimize the average ΔPSNR at the sequence level, the selection between evDA and l3DA depends on SeqDRR. For smaller SeqDRR (Fig. 4(a)), the non-reference-only allocation shows better performance. However, these SeqDRR are lower than 20%, which is not so practical in terms of efficiency for a lossy EC. evDA achieves better ΔPSNR for larger and more practical SeqDRR (Fig. 4(b)-(d)). That is to say, in this situation, the lossy compression plays a more important role than the error propagation for ΔPSNR, therefore the quality loss should be allocated to more B frames. This trend of lossy EC allocation is consistent with our theoretical analysis.

## 3. GOP-BASED DRR ALLOCATION

To reduce video quality loss through optimizing the DRR allocation on different levels in GOP structure, the overall flow is shown in Fig. 5, including two stages of training and testing. The first 8 frames of each sequence are used for training. Based on the analysis of these frames, we try to find a DRR allocation with the minimum ΔPSNR for each target SeqDRR, which is further divided into three steps. ΔPSNR for all possible DRR allocations in three levels are first evaluated. Then under a certain SeqDRR, the allocation with minimum ΔPSNR is searched on these obtained data in the first step, and viewed as the optimal DRR allocation (opDA). Finally, under different quantization parameters (Qp), these opDAs can be simplified and combined by piecewise linear fitting (defined as fopDA), which will be presented in Section 4.2. Moreover, to verify the performance for these allocations (opDA, fopDA, evDA and l3DA), they are tested on more frames, and a comparison will be given in Section 4.3.

In the training stage, all possible cases of DRR allocation are tested under the experimental condition as given in Section 4.1. DRR of each frame is fixed and ranges from 0% to 70% with a step of 2%. All frames belonging to the same level are compressed with the same DRR, defined as DRR$_{level}$. Fig. 6(a)~(c) show the distribution of ΔPSNR for different allocations of DRR$_{level2}$ and DRR$_{level3}$ when DRR$_{level1}$ is equal to 20%, 40% and 60%, respectively.

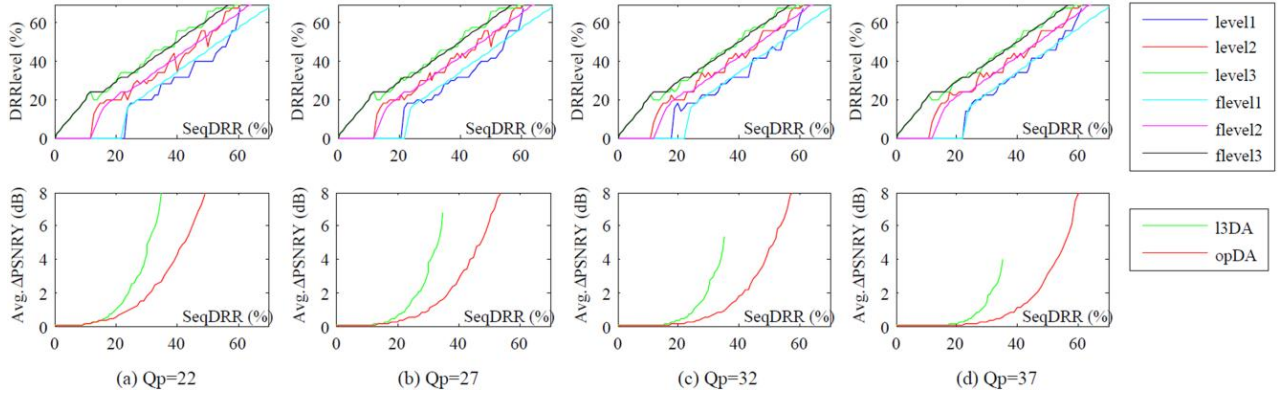

**Fig. 9.** opDA for 4 Qp values and video quality comparison. For each Qp, these two figures show the optimal DRR allocation in three levels (DRR$_{level1/2/3}$) and the average ΔPSNR respectively. "flevel" is for fopDA, which is opDA after piecewise linear fitting.

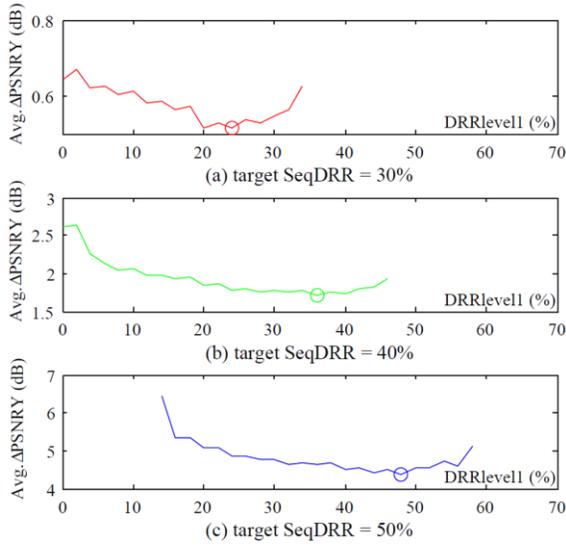

**Fig. 7.** The minimum ΔPSNR on DRR$_{level2}$ and DRR$_{level3}$ (minΔPSNR$_{23}$) from Fig. 6, and the decision of optimal DRR$_{level}$ allocation (opDA, the node in each figure) in level1.

The isoDRR line describes all the DRR$_{level}$ allocation with the same target SeqDRR. As an example, SeqDRR = 40% is plotted in Fig. 6. It shows these values of ΔPSNR related to DRR allocation in three levels under a certain SeqDRR.

To decide the opDA, the average ΔPSNR is utilized as the cost function. And it can be obtained by searching minimum value of average ΔPSNR on three variables of DRR$_{level1}$, DRR$_{level2}$, and DRR$_{level3}$. So this is an optimization on the 4-dimensional space.

To present this 4D optimization clearly, it is divided into two portions shown in Fig. 6 and Fig. 7. At first, under a target SeqDRR and a certain DRR$_{level1}$, the minimum ΔPSNR on DRR$_{level2}$ and DRR$_{level3}$ (minΔPSNR$_{23}$) is marked as the yellow nodes drawn in Fig. 6. Then for different target SeqDRR, these obtained minΔPSNR$_{23}$ with the variable of DRR$_{level1}$ are plotted together as shown in Fig. 7. Finally, the DRR$_{level}$ allocation with minimum ΔPSNR is located as the optimal DRR$_{level}$ allocation (opDA). From these two figures, the obtained opDA is related to the target SeqDRR.

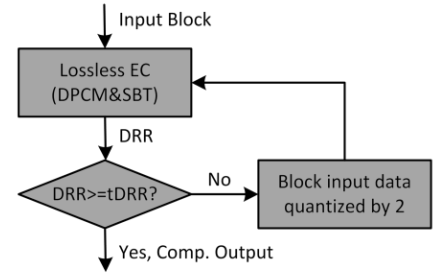

**Fig. 8.** Lossy embedded compression algorithm.

## 4. EXPERIMENTAL RESULTS

### 4.1. Experimental Conditions

A lossy EC algorithm processed in the decoder side is integrated into HM 16.0. As shown in Fig. 3, a typical random access configuration is chosen with a GOP size of 8 and an intra period of 32. The first 8 frames of 18 sequences from five classes are decoded for training. Four Qps of 22, 27, 32 and 37 are tested, but only the experimental result of Qp 32 is shown as an example in the above sections. Only the PSNR of luma component is measured.

Fig. 8 describes the lossy EC algorithm. After frames are divided into 8x8 blocks, differential pulse-code modulation (DPCM) scanning is first performed between neighboring pixels. These obtained residuals are then grouped (7 residuals/group) and coded by significant bit truncation (SBT) [1]. If DRR under the above lossless EC is smaller than the target DRR, all pixel values will be quantized until achieving the target DRR. Finally, these coded residuals are output as the compressed data before being stored into the DRAM.

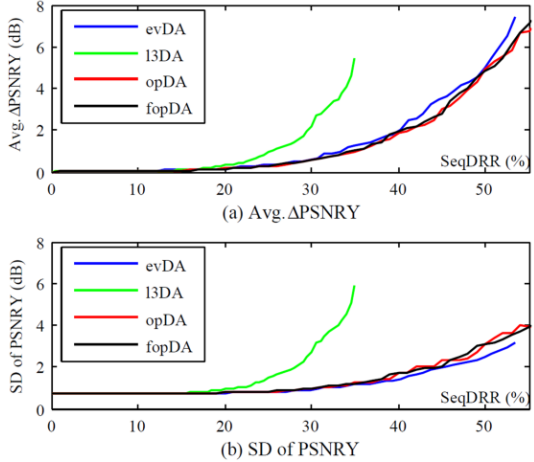

**Fig. 10.** Test and comparison of video quality loss with 100 frames.

### 4.2. Optimal DRR$_{level}$ Allocation and Comparison

Fig. 9 shows opDA for different target SeqDRR and the corresponding measurement of video quality degradation. The precision for SeqDRR is 1%, while it is 2% for DRR$_{level}$ in the range from 0% to 70%. Under a certain SeqDRR, the selected optimal DRR$_{level}$ becomes larger for higher levels.

In addition, the overall trend of optimal DRR$_{level}$ allocations are almost the same for four Qps, so they can be further simplified by piecewise linear fitting as fopDA (flevel1/2/3) shown in Fig. 9. After fitting, the obtained DRR allocation is generally suitable for all Qp values.

### 4.3. Test of DRR$_{level}$ Allocation

The optimal allocations (opDA and fopDA) for Qp 32 are tested with 100 frames. Fig. 10 shows the performance evaluation, which is similar to the analysis of opDA shown in Fig. 9(c). Except for the average ΔPSNR, the standard deviation of PSNR is also calculated to measure the fluctuation of frame quality.

The performance of fopDA is substantially the same with opDA. Therefore, with a uniform DRR$_{level}$ allocation for all Qp values, fopDA can be viewed as the best solution for DRR$_{level}$ allocation.

As a comparison, evDA and l3DA are also tested, and their results are shown in Fig. 10. When SeqDRR is larger than 17%, l3DA leads to the worst average ΔPSNR and SD of PSNR. So video quality can't be improved by reducing error propagation unilaterally. The opDA and fopDA outperform evDA on the average ΔPSNR. For some target SeqDRRs, this difference between opDA/fopDA and evDA can be up to 0.9dB. However, using optimal allocation, the SD of PSNR is a little larger than evDA, which means larger fluctuation of PSNR between frames. The average PSNR will impact more on video quality than a small fluctuation. Hence, the overall quality of opDA and fopDA should be still regarded as superior to evDA.

## 5. CONCLUSION

For the lossy EC in video decoder, this paper analyzed the influence of DRR allocation at the frame level to video quality. To achieve a target SeqDRR, it is a tradeoff of quality loss caused by the lossy compression in current frame and the error propagation. According to our experiments, completely avoiding error propagation is not effective for improving video quality.

In this paper, DPCM-based lossy EC is view as an example for testing, which is in spatial domain. So other transform-based lossy EC will be evaluated in our future study. Moreover, this analysis of frame-level DRR allocation will be extended to the low delay configuration.

## 6. ACKNOWLEDGMENT

This work is supported by Waseda University Graduate Program for Embodiment Informatics (FY2013-FY2019).